\documentstyle[aps,twocolumn,epsf]{revtex}
\begin{document}
\wideabs{
\title{Single-particle excitations and the order parameter for a trapped superfluid
Fermi gas.}
\author{M.A. Baranov}
\date{\today}
\address{FOM Institute for Atomic and Molecular Physics, Kruislaan 407,
1098 SJ Amsterdam, The Netherlands\\
Russian Research Center Kurchatov Institute, Kurchatov Square, 123182
Moscow, Russia}
\maketitle

\begin{abstract}
We reveal a strong influence of a superfluid phase
transition on the character of excitations of a trapped neutral-atom Fermi gas. 
Below the transition
temperature the presence of a spatially inhomogeneous order parameter (gap)
shifts up the excitation eigenenergies and leads to the appearance of in-gap
excitations localized in the outer part of the gas sample. The eigenenergies
become sensitive to the gas temperature and are no longer multiples of the
trap frequencies. These features should manifest themselves in a strong
change of the density oscillations induced by modulations of the trap
frequencies and can be used for identifying the superfluid phase transition.
\end{abstract}
}

\vspace{0.4cm}

\narrowtext
Physics of ultracold trapped atomic gases has attracted a lot of attention
after the discovery of Bose-Einstein condensation \cite{Cor95,Hul95,Ket95}.
As has been proven in experiments with Bose gases, the presence of a
macroscopically occupied quantum state (condensate) makes their behavior
drastically different from that of ordinary gas samples. Trapped
neutral-atom Fermi gases, being cooled to a sufficiently low temperature,
should also exhibit prominent macroscopic quantum phenomena which are mostly
related to a superfluid pairing phase transition: Below the transition point
the behavior of the gas is governed by the presence of a macroscopic wave
function of Cooper pairs (the order parameter). Possible versions of this
phase transition in atomic samples have recently been discussed in Refs. 
\cite{HS,BKK,S,BP}. However, as only a small fraction of particles is
influenced by the pairing, it is not entirely clear how the transition will
manifest itself in dynamic and kinetic properties of the gas.

In this Letter we indicate a clear way of identifying the pairing transition
in a trapped Fermi gas. We study single-particle excitations and show that
they are strongly influenced by the pairing. Above the transition
temperature $T_{c}$ the excitation eigenfrequencies are multiples of the
trap frequencies. It turns out that below $T_{c}$ the presence of the order
parameter (spatially inhomogeneous gap) shifts up the eigenenergies and ensures
the existence of in-gap excitations localized in the outer part of the gas sample
in the well formed by the gap and the trapping potential.
Due to the temperature dependence of the order parameter, the eigenenergies become 
sensitive to the gas temperature and are no longer multiplies of the trap
frequencies. This should strongly change the response of the
gas to modulations of the trap frequencies (see below). 

The influence of pairing on single-particle excitations is essentially the
same for all types of pairing discussed in 
Refs. \cite{HS,BKK,S,BP}, and for simplicity we confine ourselves to the case of
the $s$-wave pairing. We 
consider a two-component neutral gas of fermionic atoms trapped in a
spherically symmetric harmonic potential. The two (hyperfine) components
labeled as $\alpha $ and $\beta $ are assumed to have equal concentrations.
The Hamiltonian of the system has the form ($\hbar =1$)

\begin{equation}
H=\sum_{i=\alpha ,\beta }\int d{\bf r}\psi _{i}^{+}H_{0}\psi _{i}+V\int d%
{\bf r}\psi _{\alpha }^{+}\psi _{\alpha }\psi _{\beta }^{+}\psi _{\beta },
\label{1}
\end{equation}
where $\psi _{i}({\bf r})$ with $i=\alpha ,\beta $ are the field operators
of the $\alpha $ and $\beta $ atoms, $H_{0}=-\nabla ^{2}/2m+m\Omega
^{2}r^{2}/2-\mu $, $\Omega $ is the trap frequency, and $\mu $ is the
chemical potential which greatly exceeds $\Omega $ in the Thomas-Fermi limit
(see e.g. Ref. \cite{R}) discussed below. The second term in Eq. (\ref{1})
assumes an attractive elastic interaction between atoms in the states $%
\alpha $ and $\beta $ ($s$-wave scattering length $a<0$), $V=4\pi a/m$ being
the coupling constant, and $m$ the atom mass. The interaction between atoms
in the same hyperfine states, originating in the case of fermions only from
the scattering with orbital angular momenta $l\geq 1$, is here neglected.

The presence of a negative $s$-wave scattering length for the
inter-component interaction leads to a superfluid phase transition via
Cooper pairing in the $s$-wave channel \cite{S}, with the critical
temperature $T_{c}\ll \mu $. Being interested in the effect of this
transition, we consider temperatures $T\ll \mu $, where the chemical
potential coincides with the Fermi energy in the center of the trap: $\mu
\approx \varepsilon _{F}=p_{F}^{2}/2m$, and the Thomas-Fermi radius of the
gas sample $R_{TF}=v_{F}/\Omega $ serves as a unit of length ($v_{F}=p_{F}/m$%
). In this temperature range a small parameter of the theory is $\lambda
=2\left| a\right| p_{F}/\pi \ll 1$. The density profile of the gas is $%
n(R)=n_{0}(1-R^{2})^{3/2}$, where $n_{0}=p_{F}^{3}/3\pi ^{2}$ is the maximum
gas density, and $R$ the distance from the origin in units of $R_{TF}$. This
density profile corresponds to the local Fermi momentum $%
p_{F}(R)=p_{F}(1-R^{2})^{1/2}$, and the density of states on the local Fermi
surface $N(R)=mp_{F}(R)/(2\pi ^{2})$.

We assume that the critical temperature $T_{c}$ of the pairing transition is
much larger than $\Omega $ and, hence, $T_{c}$ is very close \cite{BP} to
the critical temperature $T_{c}^{(0)}=0.28\varepsilon _{F}\exp (-1/\lambda )$
in a spatially homogeneous gas with density $n_{0}$\cite{GM-B}. Below $T_{c}$
the gas is characterized by the presence of the order parameter $\Delta (%
{\bf R})=\left| V\right| \left\langle \psi _{\alpha }({\bf R})\psi _{\beta }(%
{\bf R})\right\rangle $. Following a standard mean-field procedure (see,
e.g. \cite{dG}), the term describing the interparticle interaction in Eq. (%
\ref{1}) can be written as $\Delta ({\bf R})\psi _{\alpha }({\bf R})\psi
_{\beta }({\bf R})+\Delta ^{*}({\bf R})\psi _{\beta }^{\dagger }({\bf R}%
)\psi _{\alpha }^{\dagger }({\bf R})$. Then the Hamiltonian (\ref{1})
becomes bilinear and can be reduced to a diagonal form by using the
Bogolyubov transformation generalized for the spatially inhomogeneous case 
\cite{dG}:

\[
\left( 
\begin{array}{l}
\psi _{\alpha }({\bf R}) \\ 
\psi _{\beta }({\bf R})
\end{array}
\right) =\sum_{\nu }\left[ U_{\nu }({\bf R})\left( 
\begin{array}{l}
\alpha _{\nu } \\ 
\beta _{\nu }
\end{array}
\right) +V_{\nu }^{*}({\bf R})\left( 
\begin{array}{l}
\beta _{\nu }^{\dagger } \\ 
-\alpha _{\nu }^{\dagger }
\end{array}
\right) \right] , 
\]
where $\alpha _{\nu }$ and $\beta _{\nu }$ are the operators of
single-particle excitations. Their wave functions $U_{\nu }({\bf R})$, $%
V_{\nu }({\bf R})$ satisfy the Bogolyubov-de Gennes equations

\begin{equation}
H_{0}\left( 
\begin{array}{l}
U_{\nu } \\ 
V_{\nu }
\end{array}
\right) +\left( 
\begin{array}{l}
\Delta ({\bf R})V_{\nu } \\ 
-\Delta ^{*}({\bf R})U_{\nu }
\end{array}
\right) =\varepsilon _{\nu }\left( 
\begin{array}{l}
U_{\nu } \\ 
-V_{\nu }
\end{array}
\right) ,  \label{1111}
\end{equation}
with $\varepsilon _{\nu }\geq 0$ being the excitation energies. In this
paper we develop a method of finding the order parameter $\Delta ({\bf R})$
and solving equations (\ref{1111}).

In the vicinity of $T_{c}$ the order parameter can be found from the
Ginzburg-Landau equation \cite{BP}:

\begin{equation}
\Delta (R)=5.15\cdot T_{c}\sqrt{(T_{c}-T)/T_{c}}\exp \left(
-R^{2}/2l_{\Delta }^{2}\right) ,  \label{2}
\end{equation}
where $l_{\Delta }^{2}=\kappa \sqrt{2\lambda /(1+2\lambda )}\ll 1$, and $%
\kappa =0.13(\Omega /T_{c})$. At lower temperatures, where the
Ginzburg-Landau approach is not valid, $\Delta (R)$ can be obtained from the
Eilenberger equations \cite{Eilenb} which in the presence of a harmonic
trapping potential read

\begin{eqnarray}
ig_{\omega }^{\prime }+(\Delta \cdot f_{\omega }-\Delta ^{*}\cdot \widetilde{%
f}_{\omega })/\Omega &=&0  \nonumber \\
if_{\omega }^{\prime }-(2i\omega f_{\omega }+2\Delta ^{*}\cdot g_{\omega
})/\Omega &=&0  \label{3} \\
i\widetilde{f^{\prime }}_{\omega }+(2i\omega \widetilde{f}_{\omega }+2\Delta
\cdot g_{\omega })/\Omega &=&0,  \nonumber
\end{eqnarray}
with the symbol prime standing for the operator $(p_{F}(R)/p_{F})({\bf %
n\nabla }_{{\bf R}})$. Eqs. (\ref{3}) follow from the well-known Gor'kov
equations \cite{AGD} under the assumption $\varepsilon _{F}(R)\gg T$. The
functions $f_{\omega }$, $\widetilde{f}_{\omega }$ are related to the
anomalous Matsubara Green functions in the frequency-coordinate
representation, $F=\left\langle T_{\tau }\psi ^{+}\psi ^{+}\right\rangle $
and $\widetilde{F}=\left\langle T_{\tau }\psi \psi \right\rangle $, and the
function $g_{\omega }$ to the combination of the normal Green functions, $%
\widetilde{G}=-\left\langle T_{\tau }\psi \psi ^{+}\right\rangle
-\left\langle T_{\tau }\psi ^{+}\psi \right\rangle $. In particular,

\[
f_{\omega }({\bf R},{\bf n})=\int d\xi F_{\omega }\left( {\bf R}%
,(p_{F}(R)+m\xi /p_{F}(R)){\bf n}\right) , 
\]
where $F_{\omega }({\bf R},p{\bf n})=\int d{\bf r}\exp {\bf (-}ip{\bf nr)}%
F_{\omega }({\bf R+r}/2,{\bf R}-{\bf r}/2)$ is the Fourier transform of the
anomalous Green function $F_{\omega }({\bf r}_{1},{\bf r}_{2})$ with regard
to ${\bf r=r}_{1}-{\bf r}_{2}$, and ${\bf n}$ is a unit vector. Similar
relations are valid for $\widetilde{f}_{\omega }$ and $g_{\omega }$. The
functions $g_{\omega }$, $f_{\omega }$ and $\widetilde{f}_{\omega }$ obey
the constraint \cite{Eilenb}

\begin{equation}
f_{\omega }({\bf R},{\bf n})\widetilde{f}_{\omega }({\bf R},{\bf n}%
)-g_{\omega }^{2}({\bf R},{\bf n})=1/4  \label{6}
\end{equation}
which also provides a normalization condition. Eqs. (\ref{3}) should be
completed by the self-consistency condition

\begin{equation}
\Delta (R)=\left| V\right| \cdot 2\pi N(R)T\sum_{\omega }\int \frac{d{\bf n}%
}{4\pi }f_{\omega }({\bf R},{\bf n}),  \label{4}
\end{equation}
where the summation is performed over the Matsubara frequencies $\omega =\pi
T(2n+1)$, with $n$ being an integer.

For real $\Delta $ the function $\widetilde{f}_{\omega }({\bf R},{\bf n}%
)=f_{\omega }({\bf R},-{\bf n})$, and the solution of Eqs. (\ref{3}) for $%
f_{\omega }$, omitting the terms of order $\left( \Omega /\sqrt{T^{2}+\Delta
^{2}}\right) ^{3}$, can be written as

\begin{eqnarray} 
f_{\omega } &=&\frac{\Delta }{2\sqrt{\omega ^{2}+\Delta ^{2}}}+\frac{\omega  
}{4}\cdot \frac{\Omega \Delta ^{\prime }}{(\omega ^{2}+\Delta ^{2})^{3/2}}  
\nonumber \\ 
&&+\Omega ^{2}\left[ \frac{\omega ^{2}}{16}\cdot \frac{2(\omega ^{2}+\Delta 
^{2})\Delta ^{\prime \prime }-5\Delta (\Delta ^{\prime })^{2}}{(\omega 
^{2}+\Delta ^{2})^{7/2}}\right] ,  \label{7} 
\end{eqnarray} 
where the symbol prime has the same meaning as before. Eqs. (\ref{7}) and (%
\ref{4}) provide us with an equation for the order parameter $\Delta (R)$.
Actually, the first term of Eq. (\ref{7}) gives a formally divergent
quantity $\sum_{\omega }\pi T/\sqrt{\omega ^{2}+\Delta ^{2}}$. To eliminate
this divergency we renormalize the coupling constant $V$ in the same way as
it has been done in the Bogolyubov method (see, e.g., \cite{LL}). Then we
obtain

\[ 
\frac{\Delta }{\lambda N(R)}=\Delta \cdot \widetilde{S}_{1/2}+S_{5/2}\frac{%
1-R^{2}}{12}\left[ \frac{d^{2}\Delta }{dR^{2}}\right.  
\] 
 
\begin{equation} 
+\left. \frac{1}{R}\frac{d\Delta }{dR}\cdot \frac{2-3R^{2}}{1-R^{2}}\right] 
-S_{7/2}\frac{5(1-R^{2})}{24\Omega ^{2}}\left( \frac{d\Delta }{dR}\right) 
^{2}\Delta ,  \label{8} 
\end{equation} 
where $S_{\alpha }\equiv \pi T\sum_{\omega }\omega ^{2}/(\omega ^{2}+\Delta
^{2})^{\alpha }$ for $\alpha =5/2,7/2$ , and

\[
\widetilde{S}_{1/2}=\frac{1}{\lambda }-\gamma -\ln \frac{\Delta }{\pi
T_{c}^{(0)}(1-R^{2})}-\int_{0}^{\infty }\!\!\frac{2dx}{\exp (\frac{\Delta }{T%
}\cosh (x))+1}, 
\]
with $\gamma =0.577$ being the Euler constant.

At $T$ close to $T_{c}$ equation (\ref{8}) reduces to the Ginzburg-Landau
equation from Ref. \cite{BP}. For lower temperatures we solved Eq. (\ref{8})
numerically, with the boundary conditions $\Delta ^{\prime }(0)=0$ and $%
\Delta (1)=0$. In Fig. 1 we present $\Delta (R)$ at various temperatures for 
$\lambda =0.3$ and $T_{c}^{(0)}=5\Omega $ ($T_{c}=0.86T_{c}^{(0)}$). The
comparison with results of the local-density approximation shows that the
latter is not valid in general, but gives reasonable results only at very
low temperatures: With decreasing $T$, the spatial region where $\Delta $ is
essentially nonzero increases and, hence, spatial derivatives of $\Delta $,
neglected in the local-density approximation, become less important.

\begin{figure}
\epsfxsize=8cm
\epsfbox{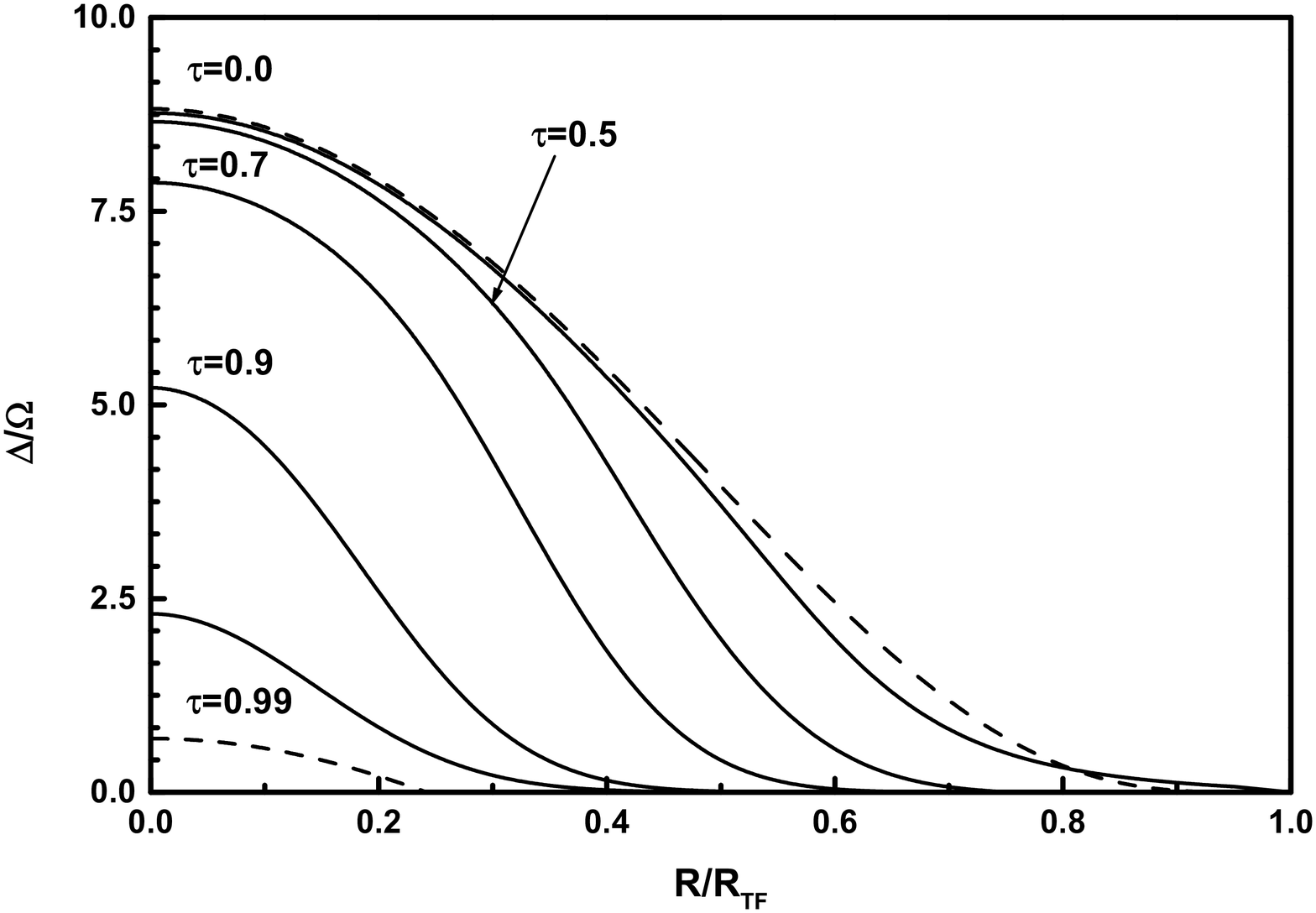}
\caption{Order parameter $\Delta (R)$ at various temperatures $\tau =T/T_{c}$
(solid lines). The dashed lines correspond to $\Delta (R)$ in the local
density approximation at $\tau =0$ (upper curve) and $\tau =0.99$ (lower
curve). }
\end{figure}

In a spherically symmetric trapping potential elementary excitations are
characterized by the radial quantum number $n$, orbital angular momentum $l$%
, and its projection $m$. The excitation wave functions can be written as $%
(U_{\nu },V_{\nu })=R^{-1}Y_{lm}(\widehat{{\bf R}})(u_{nl}(R),v_{nl}(R))$,
and the Bogolyubov-de Gennes equations (\ref{1111}) take the form

\[ 
\left[ \frac{d^{2}}{dR^{2}}-\widetilde{\mu }^{2}\left( R^{2}-1+\frac{l(l+1)}{%
\widetilde{\mu }^{2}R^{2}}\right) \right] \left(  
\begin{array}{l} 
u_{nl} \\  
v_{nl} 
\end{array} 
\right)  
\] 
 
\begin{equation} 
-2\widetilde{\mu }\widehat{\Delta }\left(  
\begin{array}{l} 
v_{nl} \\  
-u_{nl} 
\end{array} 
\right) =2\widetilde{\mu }\widehat{\varepsilon }_{nl}\left(  
\begin{array}{l} 
u_{nl} \\  
-v_{nl} 
\end{array} 
\right) ,  \label{13} 
\end{equation} 
where $\widetilde{\mu }=(2\mu /\Omega )\gg 1$, $\widehat{\varepsilon }%
_{nl}=\varepsilon _{nl}/\Omega \geq 0$, $\widehat{\Delta }=\Delta /\Omega $,
and the functions $(u,v)$ are normalized by the condition $\int_{0}^{\infty
}\left( u_{nl}u_{n^{\prime }l}^{*}+v_{nl}v_{n^{\prime }l}^{*}\right)
dR=\delta _{nn^{\prime }}$. At temperatures $T\ll \varepsilon _{F}\approx
\mu $ elementary excitations formed by particles with energies in a narrow
vicinity of the Fermi surface are most important. In the classically
accessible region of space the excitation wave functions exhibit strong
spatial oscillations, with a period of order $p_{F}^{-1}(R)\ll R_{TF}$ and a
slowly varying amplitude $\widetilde{u}_{nl}(R)$, $\widetilde{v}_{nl}(R)$:

\begin{equation}
\left( 
\begin{array}{l}
u_{nl} \\ 
v_{nl}
\end{array}
\right) =\frac{\exp \left( i\widetilde{\mu }\int_{R_{1}}^{R}p_{Fl}dR\right) 
}{\sqrt{p_{Fl}(R)}}\left( 
\begin{array}{l}
\widetilde{u}_{nl} \\ 
\widetilde{v}_{nl}
\end{array}
\right) +{\rm H.c.}.  \label{13'}
\end{equation}
The partial Fermi momentum is defined as $p_{Fl}(R)=(1-R^{2}-(l+1/2)^{2}/%
\widetilde{\mu }^{2}R^{2})^{1/2}$, and the classically accessible region is
specified by the condition $R_{1}<R<R_{2}$, with the turning points $R_{1,2}$
from the equation $p_{Fl}(R_{1,2})=0$. Omitting the terms of order $%
\widetilde{\mu }^{-1}$ in Eq. (\ref{13}), for the amplitudes $f_{\pm }=%
\widetilde{u}\pm i\widetilde{v}$ we obtain a pair of decoupled equations:

\begin{equation}
\left[ -\left( p_{Fl}\frac{d}{dR}\right) ^{2}+\widehat{\Delta }^{2}\pm p_{Fl}%
\frac{d\widehat{\Delta }}{dR}-\widehat{\varepsilon }^{2}\right] f_{\pm }=0.
\label{14''}
\end{equation}

In classically inaccessible regions $0<R<R_{1}$(due to the centrifugal
barrier) and $R>R_{2}$(due to the trapping potential), Eqs. (\ref{14''})
should be modified by replacing $p_{Fl}(R)$ with $\mp i\left|
p_{Fl}(R)\right| $, respectively, to obtain decaying solutions. Above $T_{c}$
the order parameter is zero, and using a standard semiclassical procedure
one can obtain from Eqs (\ref{13})-(\ref{14''}) the well-known result $%
E_{nl}^{(0)}=(2n+l+3/2)\Omega $ for the eigenenergies counted off the bottom
of the potential well. In a spherical harmonic trap the chemical potential $%
\mu =(j+3/2)\Omega $, where $j$ is a positive integer. Accordingly, we
obtain $\varepsilon _{nl}^{(0)}=\left| 2n+l-j\right| \Omega $ for the
eigenenergies of particle-like ($2n+l\geq j$, $\widetilde{v}_{nl}=0$) and
hole-like ($2n+l\leq j$, $\widetilde{u}_{nl}=0$) single-particle excitations.

Below the transition point the appearence of the order parameter $\Delta (R)$
modifies the excitation spectrum. Just below $T_{c}$ the order parameter is
small and exists only in a small spatial region of radius $l_{\Delta }\ll 1$
(see Eq. (\ref{2})). Therefore, the presence of $\Delta (R)$ only influences
the excitations with small $l$ and slightly shifts up their eigenenergies:
The shifts will be of order $\delta =\Delta (R_{1})l_{\Delta }$, i.e. much
smaller than the maximum value $\Delta (0)$ of the spatially inhomogeneous
gap $\Delta (R)$. Hence, the lowest excitations, namely the ones with $%
\varepsilon _{nl}^{(0)}=0$ at $T>T_{c}$, become in-gap, i.e., have energies
below the top of the gap: $\varepsilon _{nl}\sim \delta \ll \Delta (0)$.
With decreasing temperature, $\Delta (R)$ rapidly grows, which increases the
number of in-gap excited states.

At temperatures well below $T_{c}$ the characteristic radius of $\Delta (R)$
becomes of order the size of the gas sample $R_{TF}$, and all relevant
excitations ($l\lesssim \widetilde{\mu }/2$) are influenced by the presence
of $\Delta (R)$. The wave functions of above-gap excitations ($\varepsilon
_{nl}>\Delta (R_{1})$) extend over the entire classically accessible region $%
R_{1}<R<R_{2}$. On the contrary, the in-gap excitations with energies $%
\varepsilon _{nl}$ well below $\Delta (R_{1})$ are essentially ''expelled''
from the center of the trap: Their wave functions are mostly localized in
the well formed in the outer part of the sample by $\Delta (R)$ and the
trapping potential (see Fig.2).

\begin{figure}
\epsfxsize=8cm
\epsfbox{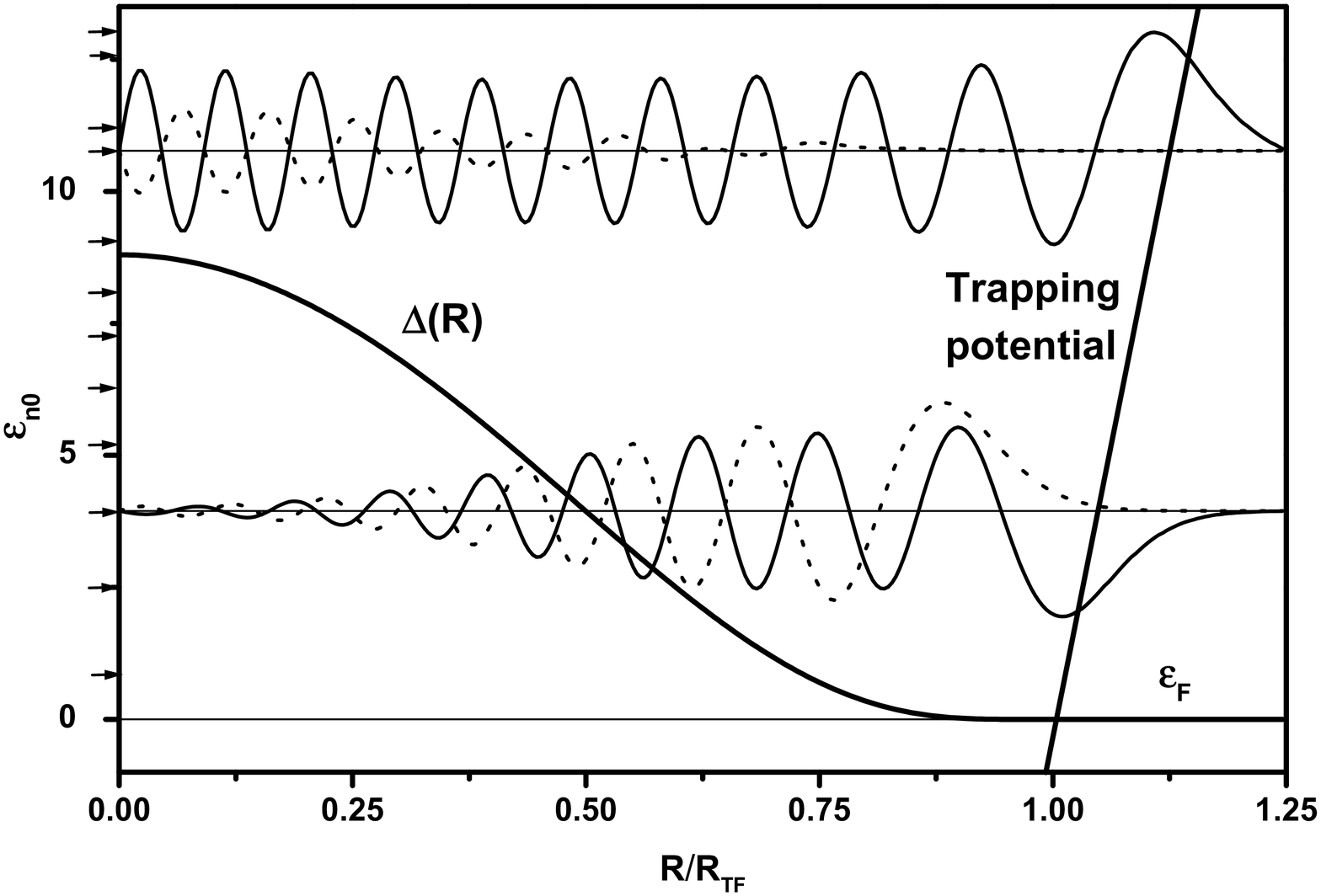}
\caption{Wave functions $u_{n0}$ (solid lines) and $v_{n0}$ (dashed lines)
for above-gap and in-gap excitations, obtained by numerical solution of Eqs.
(8) for $l=0$ at $T=0$. For illustrative purpose only (to reduce the number
of rapid oscillations) we take $\widetilde{\mu }=63$ instead of the actual
value $\widetilde{\mu }\approx 970$. Arrows
indicate eigenenergies of the excitations. }
\end{figure}

For $T$ well below $T_{c}$ we have $\widehat{\Delta }(0)\sim T_{c}/\Omega
\gg 1$, and Eqs. (\ref{14''}) for the amplitudes $\widetilde{u}_{nl}$, $%
\widetilde{v}_{nl}$ can be solved in the WKB approach, omitting the terms
with spatial derivatives of $\widehat{\Delta }(R)$ as they are small
compared to $\widehat{\Delta }^{2}$. Then, in the spatial region where $%
\varepsilon >\Delta (R)$ we obtain

\begin{eqnarray} 
\left(  
\begin{array}{l} 
\widetilde{u}_{nl} \\  
\widetilde{v}_{nl} 
\end{array} 
\right) &=&\widehat{\Delta }(R)\sum_{\pm }C_{\pm }\left(  
\begin{array}{l} 
\pm (\widehat{\varepsilon }\mp \omega (R))^{-1/2} \\  
\mp (\widehat{\varepsilon }\pm \omega (R))^{-1/2} 
\end{array} 
\right)  \label{15} \\ 
&&\times \exp \left\{ \pm i\int \frac{\omega (R)}{p_{Fl}(R)}dR\right\} .  
\nonumber 
\end{eqnarray} 
Here $\omega (R)=\sqrt{\widehat{\varepsilon }^{2}-\widehat{\Delta }^{2}(R)}$%
, and $C_{\pm }$ are numerical coefficients. In the case of above-gap
excitations ($\varepsilon _{nl}>\Delta (R_{1})$) these coefficients can be
found by making an analytical continuation of the solution (\ref{15}) to the
classically inaccessible regions $R<R_{1}$ and $R>R_{2}$. Since for $%
\varepsilon _{nl}>\Delta (R_{1})$ there are only two (classical) turning
points $R_{1}$ and $R_{2}$, the WKB quantization condition reads

\begin{equation}
\frac{2}{\pi }\int_{R_{1}}^{R_{2}}\frac{\sqrt{\varepsilon _{nl}^{2}-\Delta
^{2}(R)}}{p_{Fl}(R)}dR=\varepsilon _{nl}^{(0)}.  \label{17}
\end{equation}
This condition provides us with the energy spectrum of the above-gap
excitations (see Fig. 2). The amplitudes $\widetilde{u}_{nl}$, $\widetilde{v}%
_{nl}$ of their wave functions oscillate in the entire region $R_{1}<R<R_{2}$%
, and decay in classically inaccessible regions $R<R_{1}$ and $R>R_{2}$. As
well as excitations above $T_{c}$, these excited states are doubly
degenerate, but their energies are shifted up by the presence of $\Delta (R)$%
. More accurate analysis of Eqs. (\ref{13}) shows that the degeneracy is
lifted, but the splitting is small.

For the in-gap excitations the situation is more subtle due to the
appearance of a peculiar turning point $R_{c}$ determined by the condition $%
\varepsilon _{nl}=\Delta (R_{c})$. At this point a particle undergoes the
Andreev reflection \cite{Andr} from the spatially inhomogeneous gap $\Delta
(R)$ and transforms into a hole (and vice versa). This ensures a strong
coupling between the particle-like and hole-like excitations of the same
energy. As a result, these excitations acquire a superpositional
particle-hole character and become non-degenerate, with a splitting
increasing with $\Delta (0)$.

In the spatial region $R_{c}<R<R_{2}$ the amplitudes $\widetilde{u}_{nl}$, $%
\widetilde{v}_{nl}$ are determined by Eq. (\ref{15}), with the coefficients $%
C_{\pm }$ following from an analytical continuation of Eq. (\ref{15}) to the
region $R_{1}<R<R_{c}$. In the latter region the amplitudes are given by the
same Eq. (\ref{15}) where $\omega (R)$ is replaced by $-i\left| \omega
(R)\right| $, and the coefficients $C_{\pm }$ are obtained by making an
analytical continuation to the classically inaccessible region $R<R_{1}$. As
a result, the quantization condition for the in-gap excitations, which
provides us with their energy spectrum, reads

\begin{equation}
(-1)^{j-l}\cos (2\phi )=2Z^{2}/(Z^{4}+1),  \label{18}
\end{equation}
where $Z=\sqrt{2}\exp \left\{ \int_{R_{1}}^{R_{c}}\sqrt{\Delta
^{2}(R)-\varepsilon _{nl}^{2}}/p_{Fl}(R)dR\right\} $, and $\phi
=\int_{R_{c}}^{R_{2}}\sqrt{\varepsilon _{nl}^{2}-\Delta ^{2}(R)}/p_{Fl}(R)dR$%
. The wave functions of these excitations are mainly localized in the region 
$R_{c}<R<R_{2}$, where the amplitudes $\widetilde{u}_{nl}$ and $\widetilde{v}%
_{nl}$ oscillate. For $R_{1}<R<R_{c}$ they decay exponentially. In fact,
these amplitudes behave themselves as wave functions of bound states in the
potential well formed by the trapping potential from one side and by the
order parameter $\Delta (R)$ from the other side (see Fig. 2). The existence
of such in-gap excitations in a trapped Fermi gas originates from the
spatial inhomogeneity of $\Delta (R)$. To a certain extent these excitations
are analogous to localized states in the vortex core in ordinary
superconductors \cite{CGM}.

For the lowest in-gap excitations the WKB approach for finding the
eigenenergies $\varepsilon _{nl}$ and amplitudes $\widetilde{u}_{nl}$, $%
\widetilde{v}_{nl}$ is not adequate, and one has to solve Eq. (\ref{14''})
numerically. The energies of these excitations are very sensitive to $\Delta
(R)$, and, hence, to the gas temperature. For $\Delta (R)$ in Fig. 1 at $T=0$
we find $\varepsilon _{0}=0.85\Omega $ for the lowest excitation with $l=0$
(Eq. (\ref{18}) gives $\varepsilon _{0}=1.06\Omega $). With increasing $T$
the value of $\varepsilon _{0}$ decreases ($\varepsilon _{0}=0.23\Omega $
for $T=0.99T_{c}$) and tends to zero for $T\rightarrow T_{c}$.

In conclusion, we have found the wave functions and energy spectrum of
single-particle excitations of a trapped Fermi gas and showed that they are
strongly influenced by a pairing transition. The presence of a spatially
inhomogeneous gap $\Delta (R)$ shifts up the excitation energies and leads
to the appearance of the in-gap excitations with wave functions expelled
from the center of the trap. Due to the growth of the gap with decreasing
temperature, the excitation eigenfrequencies become quite sensitive to the
value of $T$ and are no longer multiples of the trap frequencies. These
features should lead to a strong change of the density oscillations induced
by modulations of the trap frequency, which can be used for identifying 
the pairing transition. For example, spherically symmetric modulations ($%
\delta \Omega \sim \cos (\nu t)$) cause single-particle transitions between
the states with the same orbital angular momentum $l$, and above $T_{c}$ the
amplitude of the density oscillations will exhibit resonances at frequencies 
$\nu $ which are multiples of $2\Omega $. (In this Letter we do not consider
collective excitations related to density or(and) order parameter
fluctuations. They are either overdamped or have different frequencies.)
Below $T_{c}$ the presence of $\Delta (R)$ changes the eigenfrequencies of
some of the excitations, and those will not contribute to the density
oscillations at the resonance frequencies. Hence, the resonance peaks
broaden and become smaller. For $\Delta (R)$ in Fig. 1 already at $T=0.5T_{c}
$ the eigenfrequencies of all excitations with $l\lesssim \widetilde{\mu }/2$
are altered by the pairing. Thus, the resonances in the $\nu $%
-dependence of the density oscillations, characteristic for the gas above
$T_c$, will be smeared out.
This feature will not be influenced by collective excitations related to 
fluctuations of the order parameter and not discussed above, since the 
eigenfrequencies of these excitations should be different from multiplies
of $2\Omega$.   

I acknowledge fruitful discussions at the Quantum Collectief seminar in
Amsterdam, and helpful conversations with G.V. Shlyapnikov, J.T.M. Walraven
and D.S. Petrov. This work was supported by the Stichting voor Fundamenteel
Onderzoek der Materie (FOM), by INTAS, and by the Russian Foundation for
Basic Studies (grant 97-02-16532).

\end{document}